\documentclass[preprint]{aastex}
\usepackage{psfig}
\def\HI {H\kern0.1em{\sc i}} 
\def\radm {rad m$^{-2}$} 
\def\etal   {{\rm et~al.\ }}
\def\Msunpyr{M$_\odot$ y$^{-1}$}

\begin{document}
\title{~~\\ ~~\\ Magnetic Fields in the 3C\,129 Cluster}
\author{G. B. Taylor\altaffilmark{1}, F. Govoni\altaffilmark{2}, 
S. A. Allen\altaffilmark{3}, \& 
A. C. Fabian\altaffilmark{3}}

\altaffiltext{1}{National Radio Astronomy Observatory, Socorro, NM
  87801, USA; gtaylor@nrao.edu}
\altaffiltext{2}{Istituto di Radioastronomia del CNR, Via P. Gobetti
101, I-40129 Bologna, Italy; fgovoni@ira.bo.cnr.it}
\altaffiltext{3}{Institute of Astronomy, Madingley Road, Cambridge CB3 
  0HA, UK; swa@ast.cam.ac.uk;acf@ast.cam.ac.uk}



\begin{abstract}

  We present multi-frequency VLA observations of the two radio
  galaxies 3C\,129 and 3C\,129.1 embedded in a luminous X-ray cluster.
  These radio observations reveal a substantial difference in the
  Faraday Rotation Measures (RMs) toward 3C\,129.1 at the cluster
  center and 3C\,129 at the cluster periphery.  After deriving the
  density profile from available X-ray data, we find that the RM
  structure of both radio galaxies can be fit by a tangled cluster
  magnetic field with strength 6 $\mu$G extending at least 3 core
  radii (450 kpc) from the cluster center.  The magnetic field makes
  up a small contribution to the total pressure (5\%) in the
  central regions of the cluster.  The radio morphology of 3C\,129.1
  appears disturbed on the southern side, perhaps by the higher
  pressure environment.  In contrast with earlier claims for the
  presence of a moderately strong cooling flow in the 3C\,129 cluster,
  our analysis of the X-ray data places a limit on the mass deposition
  rate from any such flow of $<$1.2 \Msunpyr.

\end{abstract}


\section{Introduction}

It has been established that radio galaxies embedded in X-ray
luminous clusters have high Faraday Rotation Measures
(RMs) ({\it e.g.}, Ge 1991; Taylor, Barton \& Ge 1994).  Of the 14
cooling flow clusters in the flux limited sample of Edge, Stewart
\& Fabian (1992)
containing a radio source stronger than 100 mJy at 5 GHz, high RMs
(greater than 1000 \radm) have been found for 9/14 (64\%) (Taylor et
al.\ 1994; Taylor et al.\ 1999).  In two with very large cooling flows
(PKS 0745$-$191 and A426) no RM can be measured due to a lack of
polarized flux.  The radio sources in two more clusters are not
co-spatial with a cooling flow, and the final source in the sample
to be properly studied is the 3C\,129 cluster.  Furthermore there
appears to be a correlation between the cooling flow rate, $\dot{M}$,
and the magnitude of the RMs (Taylor et al.\ 1994).   Lower, but
significant RMs have been observed in the A119 cluster 
(Feretti \etal\ 1999) which was initially reported to have a weak 
cooling by Edge \etal (1992), but later found to have no 
cooling flow (Cirimele, Nesci \& Trevese 1997).   Some correlation
is to be expected since $\dot{M}$ and RM are both related to the
central density.  The details of this correlation, however, are yet to
be understood.  In particular we would like to understand how the
magnetic fields scale with the density, and if the cooling flow has
somehow enhanced the cluster magnetic fields.  To this end the study
of radio galaxies embedded in clusters both with and without cooling
flows is essential.

The RM structure of extended radio galaxies embedded in 
clusters can provide information about the cluster magnetic fields
unavailable by any other means.  Besides a measurement of the magnetic
field strength (determined in conjunction with X-ray observations of
the hot gas), by studying the spatial distribution of the RMs we can
estimate the coherence length of the magnetic fields.  The RM
structure of some cluster radio sources such as Hydra~A
(Taylor \& Perley 1993), Cygnus~A (Dreher, Carilli \& Perley 1987),
and A1795 (Ge \& Owen 1993) show ordered structures up to 100 kpc in
extent.  The origin and evolution of the magnetic fields required to
produce such RM structures is a topic of some speculation (Goldman \&
Rephaeli 1991, DeYoung 1992).  They could be remnant magnetic fields
of extinct radio galaxies, the result of a cluster dynamo effect
(Ruzmaikan et al. 1989), or magnetic fields stripped from spiral
galaxies during galactic cannibalism.  Cluster magnetic fields are
likely to have important effects on cluster dynamics and energy
transport (Sarazin 1986; and Tribble 1989), and significant
implications for galaxy and cluster formation.  By mapping the RM
distribution across an embedded source we can measure the correlation
length of the cluster magnetic field, and look for sign reversals.

Clusters which contain more than one moderately strong, extended
source provide valuable information about the overall extent of the
magnetic fields within the cluster.  In a recent study of three radio
galaxies embedded in the A119 cluster, Feretti et al. (1999), found
that the RMs decreased with distance from the cluster center, but were
consistent with a magnetic field strength of 10 $\mu$G.  The 3C\,129
cluster is exceptional in that it is one of very few clusters
containing two moderately strong ($>$ 100 mJy at 5 GHz) radio sources
(see Fig.~1).  In fact in the early stages of this project one of us
(GBT) failed to appreciate the presence of 3C\,129.1 {\it at the
  center of the cluster}, so only the RM structure of 3C\,129 (which
lies 15 arcminutes or a project distance of 400 kpc to the west of the
cluster X-ray center) was originally studied.

Observations of both 3C\,129 and 3C\,129.1 are presented in \S2.
In \S3 we present a detailed multi-frequency, polarimetric study
of both 3C\,129 and 3C\,129.1 including the RM distribution for both radio
galaxies.  Using these data we attempt in \S4 to characterize the
magnetic fields (strengths, coherence length, and overall extent) in
the 3C\,129 cluster.

We assume H$_0 = 65$ km s$^{-1}$ Mpc$^{-1}$ and q$_0$=0.5 throughout.
  
\section{Radio Observations and Data Reduction}

Observations of both 3C\,129 and 3C\,129.1 were made at a total of 6
frequencies distributed across the 5 and 8 GHz bands of the VLA (Very
Large Array)\footnote{The National Radio Astronomy Observatory is a
facility of the National Science Foundation operated under cooperative
agreement by Associated Universities, Inc.} telescope in 1994 and
2000.  The details of the observations are provided in Table 1.  The
source 3C\,138 was used as the primary flux density calibrator, and as
an absolute reference for the electric vector polarization angle
(EVPA).  Phase calibration was derived from the nearby compact source
0420+417 with a cycle time between calibrators of about 12 minutes.
Instrumental calibration of the polarization leakage terms was
obtained using 0420+417 which was observed over a wide range in
parallactic angle.  The data were reduced in AIPS (Astronomical Image
Processing System), following the standard procedures.  The AIPS task
IMAGR was used with a suitable taper to make $I$, $Q$ and $U$ images at each
of the 6 frequencies observed at the same resolution.  Polarized
intensity, $P$, images and polarization angle, $\chi$, images were
derived from the $Q$ and $U$ images.  The Faraday Rotation Measure (RM)
image was derived from pixel by pixel fits to $\chi$ versus $\lambda^2$ 
for all 6 $\chi$ images for 3C\,129,
but the 4.585 GHz data had to be omitted for 3C\,129.1 since the polarimetric
data at this frequency was corrupted by an incorrect R-L delay setting
at the time of observation.  Stokes $I$ images using multiple
frequencies within an observing band were also produced. In addition
to the added sensitivity, the image produced also benefits from
improved ($u,v$) coverage.  No correction for the spectral dirty beam
has been performed, but the sidelobes expected from this effect 
should be below the thermal noise floor.

\section{X-ray Observations}

\def\apc{\rm atom cm$^{-2}$}

\subsection{ASCA constraints on the X-ray temperature}

The 3C\,129 cluster was observed with the ASCA satellite on 1998 August 31. 
For our analysis, we have used the 
screened event lists available from the Goddard Space Flight Center 
(GSFC) ASCA archive. The data were reduced using the FTOOLS software 
(version 4.2) from within the XSELECT environment (version 1.4b). 
Further data-cleaning procedures including manual screening based 
on the individual instrument light curves were followed, 
as recommended in the ASCA Data Reduction Guide 
(http://legacy.gsfc.nasa.gov/docs/asca/abc/abc.html). 

Our primary goal with the ASCA analysis was to obtain a reliable 
measurement of the mean X-ray gas temperature in the cluster. For this 
purpose, only data from the two Gas scintillation Imaging 
Spectrometers (GIS) were used, which had good exposure times 
of $4.34 \times 10^4$ and $4.32 \times 10^4$s respectively. Spectra were 
extracted from circular regions of 9 arcmin radius, centred on the peak of the 
emission in each detector. Background spectra were determined from 
the `blank sky' observations of high Galactic latitude fields compiled 
during the performance verification stage of the mission. The background 
data were screened and grade selected in an identical manner to the target 
observations and the background spectra were extracted from the same 
regions of the detectors as the target spectra. 

The modelling of the ASCA spectra has been carried out using the XSPEC
code (version 11.0; Arnaud 1996). Only counts in the energy range $1.5-10.0$ 
keV were used (the energy range over which the calibration of the GIS 
detectors is currently best understood). The spectra were grouped before 
fitting to ensure a minimum of 20 counts per PHA 
channel, thereby allowing $\chi^2$ statistics to be used.
The appropriate spectral response matrices, issued by GSFC on 1995 March 6, 
were used. Auxiliary response files were generated using the ASCAARF 
software, with effective area calculations appropriate for an extended source.

The spectra have been modelled using the plasma code of Kaastra \& Mewe
(1993; incorporating the Fe L calculations of Liedhal, Osterheld \&
Goldstein 1995) and the photoelectric absorption models of Balucinska-Church 
\& McCammon (1992). The data for both GIS detectors were analysed
simultaneously, using a model appropriate for an isothermal plasma in 
collisional equilibrium at the optically-determined redshift for the 
cluster, absorbed by an equivalent hydrogen column density, $N_{\rm H}$ of 
solar metallicity cold gas. The temperature ($kT$), metallicity ($Z$) and 
absorbing column density were linked to take the same values across the two 
GIS data sets, although the normalizations were allowed to vary as independent 
fit parameters. We find that this model provides an acceptable description 
of the ASCA data with $\chi^2=995$ for 1012 degrees of freedom and best-fit 
parameter values $kT = 6.25^{+0.27}_{-0.26}$ keV, $Z=0.31\pm0.03$ solar 
and $N_{\rm H} = 6.41\pm0.44 \times 10^{21}$ \apc~(90 per cent confidence 
limits).

The mean X-ray gas temperature measured within the central 9 arcmin radius
with ASCA is consistent with the value of $5.6^{+0.7}_{-0.6}$ keV determined 
by Edge \& Stewart (1991) using EXOSAT Medium Energy Proportional 
Counter Array observations (which covered a larger $45\times45$ arcmin$^2$ 
field of view). Our ASCA result is slightly hotter than the value of 
$5.5\pm0.2$ keV measured by Leahy \& Yin (2000) from a joint analysis of 
the same EXOSAT data and a short (500s) ROSAT Position Sensitive Proportional 
Counter observation. The absorbing column density measured with ASCA 
is in good agreement with the Galactic value along the line of sight 
to the cluster of $7.1 \times 10^{21}$ \apc~(Dickey \& Lockman 1990) or 
$5.8 \times 10^{21}$ \apc~(Stark et al. 1992).

\subsection{The X-ray gas density profile}

Pointed X-ray imaging observations of the 3C\,129 cluster have been 
made with the High Resolution Imager (HRI) and Imaging 
Proportional Counter (IPC) on the Einstein Observatory and the HRI 
on ROSAT. The observation best suited to the current task of measuring the 
X-ray gas density profile in the regions surrounding the 
3C\,129 and 3C\,129.1 radio galaxies is the Einstein IPC 
observation, which was made on 1980 October 4. For our analysis, we have used 
the processed $0.2-3.5$ keV image available from the GSFC on-line archive. The 
net exposure time for the IPC observation is 6847s.

In order to determine the X-ray gas density profile and to discriminate on 
the presence/absence of any cooling flow in the cluster core, we have 
carried out a deprojection analysis of the Einstein IPC image using an 
updated version of the deprojection code of Fabian et al. (1981; see also 
White, Jones \& Forman 1997 for details). An azimuthally-averaged X-ray 
surface brightness profile was constructed from the IPC image, with a 
radial bin-size of 0.8 arcmin (23 kpc). This profile is used as the primary 
input for the deprojection analysis. The deprojection method also requires the 
total mass profile for the cluster to be specified. We have iteratively 
determined the mass profile (parameterized using the formalism
of Navarro, Frenk \& White 1997) that results in a deprojected 
temperature profile that is approximately 
isothermal, within the regions probed by the IPC data, and is 
consistent with the mean gas temperature determined from the 
ASCA data.\footnote{The best-fit parameterized mass model has a scale 
radius $r_s \sim 0.8$ Mpc, a concentration parameter $c \sim 2.7$ and an 
equivalent velocity dispersion 
$\sigma =  \sqrt{50} H_0 r_{\rm s} c \sim 1000$ km s$^{-1}$.}

The electron density profile for the central 600 kpc ($\sim 21$ arcmin) 
radius region, determined from the deprojection analysis, 
is shown in Fig.~2. The best-fitting $\beta-$model (overlaid) has  
parameter values $r_{\rm c} = 146\pm26$ kpc, $\beta=0.47\pm0.05$ and 
$n_{\rm e}(0)=2.8\pm0.3 \times 10^{-3}$ cm$^{-3}$. We note, however, that 
the $\beta-$model provides a poor fit to the data, with a 
reduced $\chi^2$ value $\sim 3.6$. The formal errors on the 
best-fit parameter values ($\Delta \chi^2=2.7$ limits) are therefore 
likely to underestimate the true uncertainties. Our results on the density 
profile are in good agreement with those presented by Leahy \& Yin (2000).

The central cooling time in the cluster (the mean cooling time within
the central 0.8 arcmin bin) is $t_{\rm cool}=1.32^{+0.66}_{-0.36}
\times 10^{10}$ yr. We determine a 90 per cent confidence upper limit
on the cooling radius (at which the cooling time first exceeds a
Hubble time) of $r_{\rm cool} < 34$ kpc, and a limit on the mass
deposition rate within this radius of $<1.2$ \Msunpyr. These results
are consistent with the previous findings of White et al. (1997) and
show that the 3C\,129 cluster does not contain a strong cooling flow.
They are inconsistent with the results of Leahy \& Yin
(2000) who claim a cooling flow of 84 \Msunpyr\ based on the
excess luminosity over their $\beta$-profile fit.

\section{Results}

We present multifrequency radio observations, polarization 
properties, and rotation measure distributions for 3C\,129 and
3C\,129.1.  Properties derived for both sources are summarized in Table 2.

\subsection{3C\,129, Cluster Periphery}

The radio source 3C\,129 was identified with an E galaxy by Hill \&
Longair (1971), and was among the first radio galaxies discovered to have
a head-tail structure.  Since then it has been studied extensively in
the radio by several authors (e.g., Miley 1973, Perley \& Erickson
1979, van Breugel 1982, Downes 1984, Feretti \etal\ 1998).  Low
frequency observations by van Breugel \& J\"agers (1982) show that the
radio tail extends over $\sim$20\arcmin\ in length ($>$ 0.5 Mpc),
making 3C\,129 sometimes classified as one of the giant radio galaxies
(Ishwara-Chandra \& Saikia 1999).  Our relatively high frequency 
interferometry observations are only sensitive to the brighter
emission from the inner 2\arcmin\, or ``head'' of 3C\,129.

The jets in 3C\,129 start out
oppositely directed nearly north-south and then curve continuously to
the northwest.  The more prominent, northern jet expands gradually,
while the southern jet flares more abruptly where it makes a sharp
bend about 30\arcsec\ west of the nucleus.  At large distances from the
nucleus the lobes merge together and bend to the west and
southwest.  We measure an integrated flux density of 1300 mJy (49\%) out of
the total flux density of 2650 mJy measured with the Effelsberg 100 m
antenna (Feretti \etal\ 1998).  Our images agree with, but have
greater sensitivity than the early 5 GHz VLA images presented by
Rudnick \& Burns (1981).

At 8.5 GHz the polarized intensity in the jets is around 10\% and increases 
to 20-30\% in the lobes.  The core is less than 0.5\% polarized.
Similar values are seen at 5 GHz, except in some filamentry regions
of very low polarization running across the lobes.  A plot of the
polarized intensity and magnetic field vectors (corrected for
the effects of RM) is shown in Fig.~3.

Early Faraday Rotation Measure (RM) estimates for 3C\,129 were $-$30
\radm (van Breugel 1982), but further observations by Downes (1984)
demonstrated that the early observation suffered from an $n\pi$
ambiguity and the correct value (at 4.3\arcmin\ resolution) was $-$130
$\pm$ 5 \radm.  At high resolution (1.8\arcsec) we find
the RMs in 3C\,129 range from $-$400 to 100 \radm\ (see Fig.~4), but
cluster around the average value of $-$125 \radm, with a dispersion of
82 \radm.  The error in the determination of the RM varies with SNR
across each source, but is generally less than 20 \radm (1$\sigma$).
Downes (1984) also found that the fractional polarization continues to
increase along the tails to a peak of about 50\% at 10\arcmin\ from
the nucleus while the integrated RM stays constant.  These RMs are
somewhat larger than expected for a typical field galaxy along this
line-of-sight which are in the range $-$40 to 40 \radm\
(Simard-Normandin, Kronberg \& Button 1981), but substantially lower
than are found for 3C\,129.1 at the cluster center.  The RMs in Fig.~4
are patchy with a scale of $\sim$5--10 kpc.  No apparent correlation of
the RMs with total or polarized intensity is seen.

\subsection{3C\,129.1, Cluster Center}

The radio galaxy 3C\,129.1 is usually mentioned in passing in papers
concentrating on its larger and brighter cousin.  In early papers its
proximity and similar redshift to 3C\,129 was taken as indicative of
clustering, although the low galactic latitude made optical
identification of other cluster members difficult.  The discovery of
strong and extended X-ray emission by the {\it Uhuru} satellite 
confirmed that both radio galaxies
are indeed members of a cluster.

The radio source 3C\,129.1 is an
order of magnitude more compact (about 1.5\arcmin) than 3C\,129 (see
Fig.~1).  The morphology is similar to most FR~I radio galaxies, with
bright regions on either side of the core that fade with distance from
the core, but on the southwest side there is a sharp boundary to the
radio emission that is uncharacteristic of FR~Is.

At both 5 and 8 GHz, the bright inner 30\arcsec\ are polarized at the
level of 10\%.  The fractional polarization increases to 20-30\% in
the fainter outer regions.  The core has less than 2\% polarization.
A plot of the polarized intensity and magnetic field vector
orientation is shown in Fig.~5.  Examination of polarimetry from the
NVSS at 1.4 GHz (Condon \etal\ 1998) shows between 0.2 and 1\%
polarization in 3C\,129.1 compared to 1 -- 40\% in 3C\,129 --
consistent with the depolarizing effects of high RMs towards
3C\,129.1.

The RMs in 3C\,129.1 are shown in Fig.~6 and range from $-$600 to
$+$750 \radm, considerably larger than those in 3C\,129. Histograms of
the RM distributions for both sources (Figures 7 and 8) illustrate the
differences in both the average and the dispersion of the RM between
these two sources.  The error in the determination of the RM varies
with SNR across each source, but is generally less than 20 \radm
(1$\sigma$).  The scale size of the RMs in 3C\,129.1 appears smaller,
$\sim$3 kpc.  Over most of the sources the RMs trend up or down
smoothly, but there are a few sign reversals between adjacent patches.

\section{Discussion}

\subsection{Radio Source Properties}

The head-tail radio sources like 3C\,129 are usually found at the
periphery of clusters.  The usual explanation for this morphology
is that the jets are swept back by relative motion between the
radio galaxy and the IGM. Rudnick \& Burns (1981) estimated a
relative velocity between 3C\,129 and the IGM of 3000 km s$^{-1}$
using a cluster gas density of $\sim$10$^{-4}$ cm$^{-3}$.  This
is comparable to the density derived from the X-ray observations
in \S3.2 of 6 $\times$ 10$^{-4}$ cm$^{-3}$.

The compactness of 3C\,129.1 is probably a direct result of its 
central location within the cluster.  This is similar, but less
extreme than the distortions seen in the radio emission from the
cD galaxies PKS~0745$-$191 (Taylor, Barton \& Ge 1994) and
3C\,317 (Zhao \etal\ 1993).

                                        
\subsection{The Galactic RM Contribution}

At the low galactic latitude of the 3C\,129 cluster (l,b = 160.4,
0.1$^\circ$), the galactic RM contribution could be substantial.  This
contribution may be in the form of (1) an additive offset to the RMs,
and (2) point-to-point variations in the RM distribution.  In regards
to point (1) the constant RM of $-$130 \radm\ over 10\arcmin\ (12 pc
at a distance of 4 kpc, 250 kpc at the distance of 3C\,129) found by
Downes (1984) is more readily explained as the result of a Galactic
offset rather than a cluster magnetic field organized over such a
large scale.  Regarding point (2) above, Clegg et al. (1992) have
examined the RMs of extragalactic sources between galactic longitudes
of 30 and 90 degrees in order to examine the magnetoionic structure of
the Galaxy.  For sources at low galactic latitudes ($|b| < 5^\circ$)
with two or more components seperated by less than 1$^\circ$ they find
RM differences up to 180 \radm, although typical values are more like
50 \radm.  In 3C\,129 the RM difference between adjacent regions in
the radio lobes ranges up to 400 \radm.  Unless the Galaxy has an
exceptionally strong and tangled magnetoionic structure at $l =
160^\circ$, our observations of 3C\,129 are difficult to explain
entirely by a galactic RM.  For several radio galaxies within 10
degrees of the 3C\,129 cluster (including 3C\,134 at $l, b = 168,
-1.9^\circ$) Simard-Normandin \etal\ (1981) find RMs in the range
$-$40 to 40 \radm, indicating that the galactic field strengths are
not particularly large in this direction.  The nearly factor 3 larger
RM dispersion in 3C\,129.1 would be even more difficult to explain as
a galactic effect.  For the remainder of this paper we consider the
3C\,129 cluster to be the dominant contributor to the RM variations
seen across both 3C\,129 and 3C\,129.1.  This interpretation is
consistent with the accumulating evidence for cluster magnetic fields
(discussed in \S1) and the large contrast in RM distribution between
3C\,129.1 at the cluster center and 3C\,129 at the periphery.

\subsection{Cluster Magnetic Field Strengths and Topologies}

For Faraday rotation from a magnetized cluster gas, the RMs are
related to the density, $n_{\rm e}$, and magnetic field along the 
line-of-sight, $B_{\|}$, through the cluster according to
$$ RM = 812\int\limits_0^L n_{\rm e} B_{\|} {\rm d}l ~{\rm
  rad~m}^{-2}~,
\eqno(1)
$$
where $B_{\|}$ is measured in $\mu$Gauss,  $n_{\rm e}$
in cm$^{-3}$ ~and d$l$ in kpc.   If the density distribution and
magnetic field topology are known then it is possible to solve
for the magnetic field strength.  

The simplest case is that of a constant density and uniform magnetic
field out to some distance from the cluster center.  If we consider
the maximum RM from 3C129.1 at the center of 600 \radm, then using the 
central density and core radius derived in \S3.2, the
magnetic field needed to produce this RM is 1.8 $\mu$G.  
This field can be thought of
as a lower limit to the true field strength since any reversals of the
magnetic field would require a larger field strength to produce the
observed RM.  Such field reversals are certainly present as the patchy
RM distribution in 3C\,129.1 (Fig.~6) clearly indicates.

A more reasonable field topology is that of cells of constant size,
density, and magnetic field strength, but a random magnetic field
direction, uniformly filling the entire cluster.  The Faraday RM
produced by such a screen will be built up in a random walk fashion
and will thus have an average value of 0 \radm, but a dispersion in
the RM, $\sigma_{\rm RM}$, that is proportional to the square root of
the number of cells along the line-of-sight through the cluster.  For
a density distribution that follows a $\beta$-profile, Felten (1996)
has derived the following relation for the RM dispersion:

$$
\sigma_{\rm RM} = {{{\rm K} B~n_0~r_c^{1/2} l^{1/2}}
\over{(1 + r^2/r_c^2)^{(6\beta - 1)/4}}}
\sqrt{{\Gamma(3\beta - 0.5)}\over{{\Gamma(3\beta)}}}
\eqno(2)
$$
where $n_0$, $r_c$, and $\beta$ are  
derived in \S3.2, $l$ is the cell size, $r$ is the distance of
the radio source from the cluster center, $\Gamma$ is the Gamma
function, 
and K is a factor which depends on the location of the radio 
source along the line-of-sight through the cluster: K = 624
if the source is beyond the cluster, and K = 441 if the source
is halfway through the cluster.  We assume that both 3C\,129
and 3C\,129.1 are embedded halfway through the cluster.  This is
a good assumption for 3C\,129.1 at the center of the cluster,
but could be considerably off for 3C\,129 at the periphery.
For the field size, $l$, we use the scale size of 5 kpc
seen in the RM distributions.  With the above measurements 
or assumed values, we use the estimated
RM dispersion (Figs.~9 and 10) for 3C\,129 and 3C\,129.1
to estimate the cluster magnetic field strength at 6.1 and 5.6 $\mu$G 
respectively.  The good agreement between the field
strengths estimated from these two sources is encouraging,
although this model is probably still too simple to hold
up to more rigorous tests.  Probably both
$l$ and $B$ change with distance from the cluster center,
and the density distribution may be considerably more complex,
especially at the center of the cluster and in the vicinity
of a powerful radio galaxy.  

The estimated cluster magnetic field strength of 6 $\mu$G compares
favorably to the magnetic field strength of 5--10 $\mu$G found in A119
by Feretti \etal\ (1999) through a similar analysis of 3 embedded
radio galaxies.  These field strengths are somewhat less than the
$\sim$30 $\mu$G derived for sources embedded in strong cooling flows
like Hydra A (Taylor \& Perley 1993).  The magnetic field pressure in
the 3C\,129 cluster corresponds to 5\% of the combined magnetic and
thermal gas pressure at the cluster center.  At the projected distance
of 3C\,129 of 400 kpc, this simple model for the magnetic field
predicts nearly a fifth of the combined pressure comes from the magnetic
fields.

Based on the data presented here we cannot rule out the possibility
that the cluster gas density and magnetic fields are enhanced in the
vicinity of the radio sources and that this ``cocoon'' provides the
observed Faraday screen.  However, the recent demonstration of a RM excess
towards {\it background} radio sources seen through clusters by
Clarke, Kronberg \& B\"ohringer (2001) argues convincingly in favor of
cluster-wide magnetic fields.

\section{Conclusions}

The 3C\,129 cluster is exceptional in that it possesses two bright,
extended radio galaxies embedded in a hot X-ray emitting gas.  We have
reanalyzed available X-ray data to determine the density profile and
combined this information with high resolution radio measurements of
the Faraday Rotation Measure structure to provide information on the
magnetic field strengths and topologies.  The Faraday Rotation Measure
images reveal a patchy structure with a scale length of $\sim$5 kpc
indicating that the magnetic fields are tangled on this scale.  A
simple model with a magnetic field in randomly oriented cells of
uniform size and strength, and a gas density distribution given by a
$\beta$ profile, yields a magnetic field strength of $\sim$6 $\mu$G
out to $\sim$450 kpc.  These magnetic fields will have a significant
impact on energy transport through the cluster, and may be dynamically
important as well.  Although our reanalysis of the X-ray data does not
reveal the cooling flow suggested in the past (Edge \etal\ 1992; Leahy
\& Yin 2000), the field strengths derived are similar to those found
in cooling flow clusters.  This supports claims by Feretti \etal
(1999) and Clarke, Kronberg \& B\"ohringer (2001) that magnetic fields
are pervasive in galaxy clusters.

We also find that the deeply 
embedded radio galaxy 3C\,129.1 appears more strongly confined on the 
southern side.  Higher resolution X-ray observations with Chandra 
will reveal if the thermal gas pressure is enhanced in this region.

Future observations with an Expanded Very Large Array (EVLA) could
increase the number of radio galaxies for which we can image the
rotation measure distribution tenfold.  This would provide many more
constraints on the magnetic field topology in clusters.

\acknowledgments
SWA and ACF acknowledge the support of the Royal Society.
This research has made use of the NASA/IPAC Extragalactic Database (NED)
which is operated by the Jet Propulsion Laboratory, Caltech, under
contract with NASA.

\clearpage

\begin{figure}
\vspace{19cm}
\includegraphics{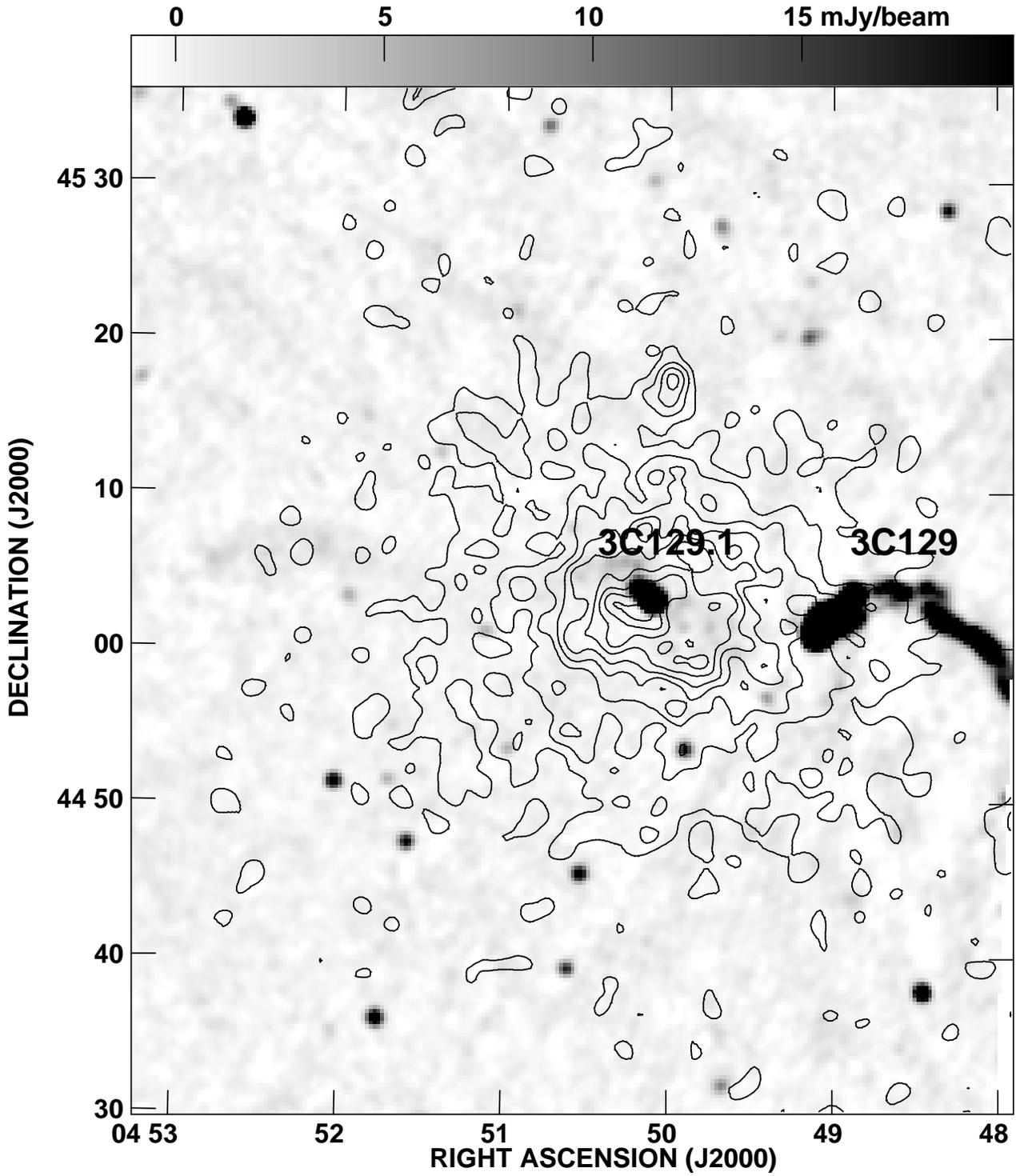}
\caption{The 3C\,129 cluster in the X-ray (contours from the
Einstein IPC image) and radio (greyscale from the NVSS, with range
$-$1 to 20 mJy/beam.).  The
resolution is roughly 45\arcsec\ in both images.
\label{fig1}}
\end{figure}
\clearpage

\begin{figure}
\vspace{19cm}
\includegraphics{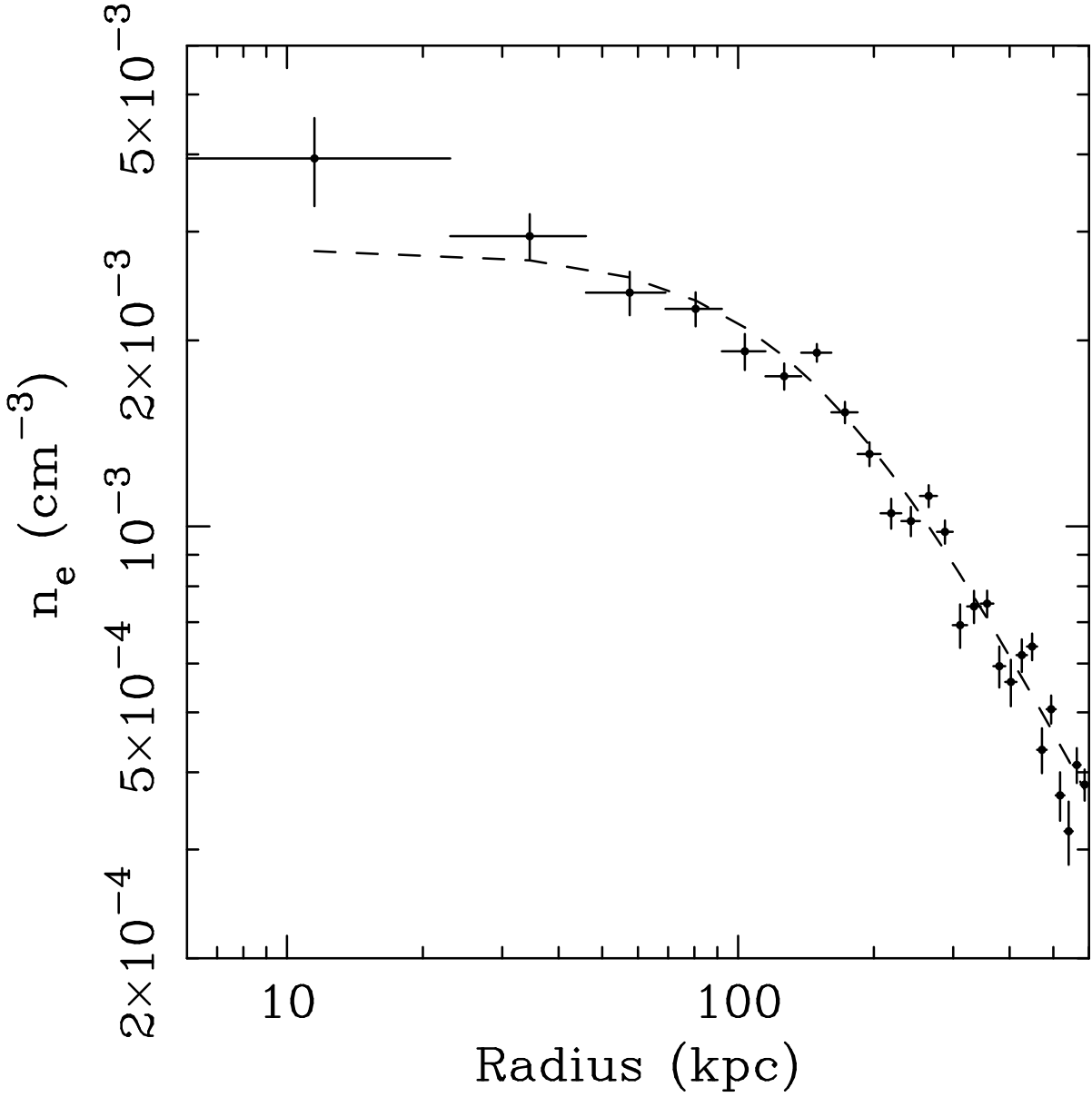}
\caption{The electron density profile determined from the 
deprojection analysis 
of the Einstein IPC X-ray image. Error bars are $1\sigma$ uncertainties 
determined from 100 Monte Carlo simulations. A mean gas temperature of 
6.25 keV, an absorbing column density of $6.4 \times 10^{21}$ \apc~and a 
metallicity of 0.31 solar are assumed (in agreement with the results 
determined from the ASCA spectra).
\label{fig2}}
\end{figure}
\clearpage


\begin{figure}
\vspace{19cm}
\includegraphics{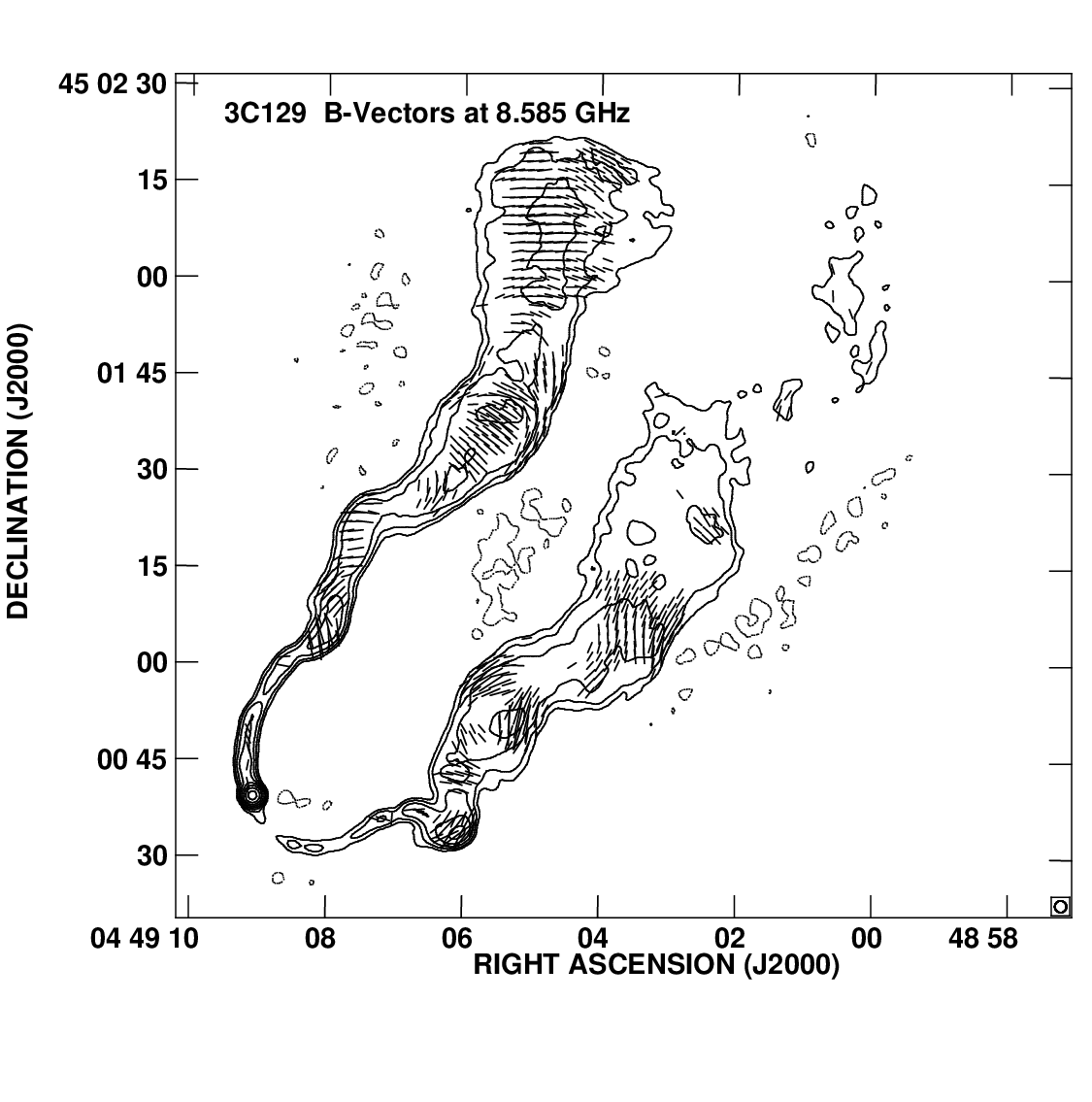}
\caption{Polarized intensity projected magnetic field vectors at 
8.6 GHz (1 arcsec = 0.14 mJy/beam; vectors corrected for Faraday rotation) 
overlaid on
total intensity contours for 3C\,129.  Contours start at 0.2 mJy/beam
and increase by factors of 2. The peak in the image is 37
mJy/beam. The restoring beam is plotted in the lower right corner and
is a circular Gaussian with a FWHM of 1.8 arcseconds.
\label{fig3}}
\end{figure}
\clearpage

\begin{figure}
\vspace{19cm}
\includegraphics{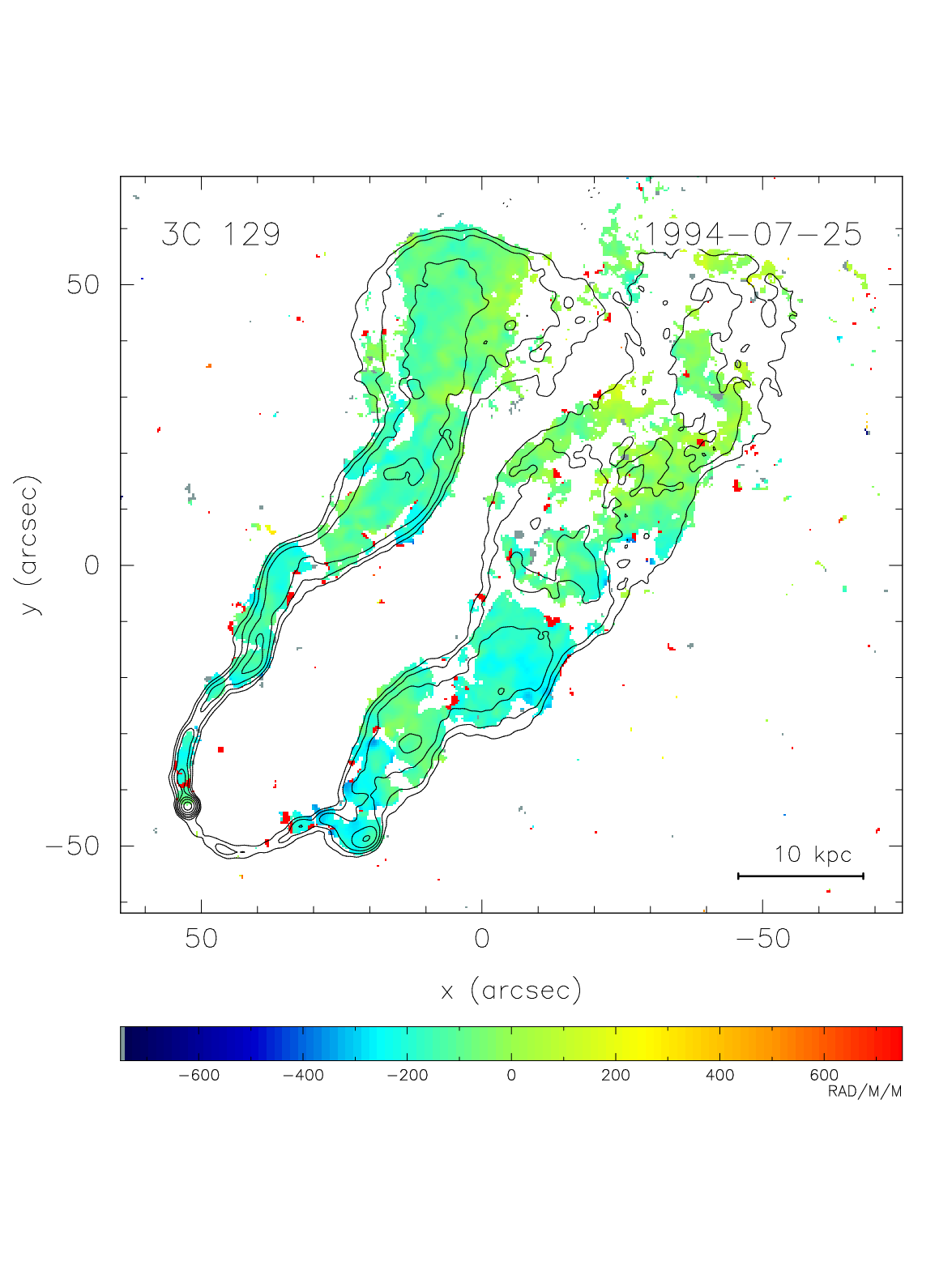}
\caption{Rotation measure image of 3C\,129 with contours of total
intensity at 4.7 GHz overlaid.  The colorbar ranges from $-$750 to $+$750
\radm.  Due to the low blanking level chosen in order to see the RMs
in the faint lobes, some spurious RMs show up as red or black spots at
the edges (or beyond) regions of polarized emission.
\label{fig4}}
\end{figure}
\clearpage


\begin{figure}
\vspace{19cm}
\includegraphics{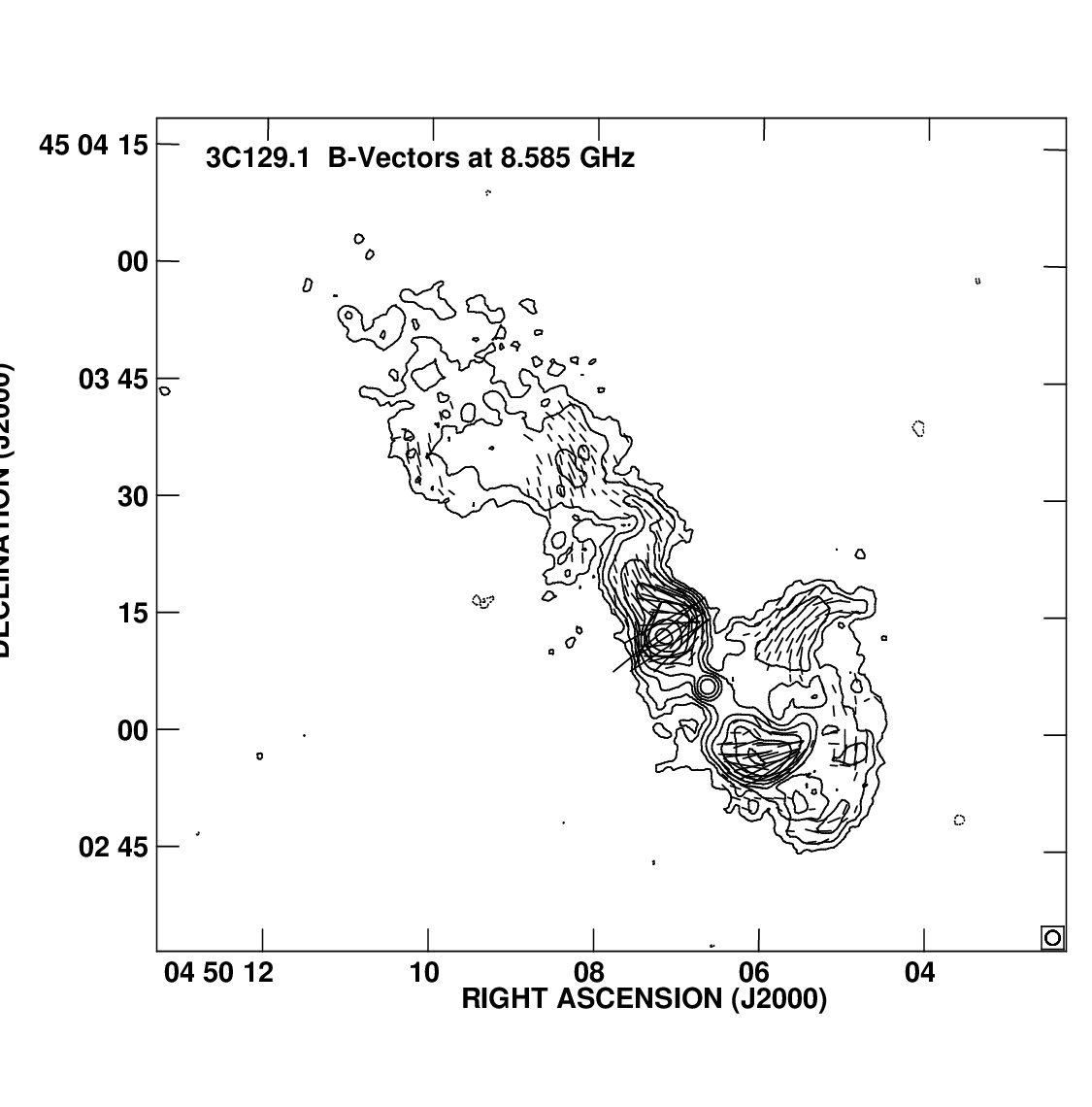}
\caption{Polarized intensity projected magnetic field vectors at 
8.6 GHz (1 arcsec = 0.071 mJy/beam; 
vectors corrected for Faraday rotation) overlaid on
total intensity contours for 3C\,129.1.  Contours start at 0.05 mJy/beam
and increase by factors of 2. The peak in the image is 7.9
mJy/beam. The restoring beam is plotted in the lower right corner and
is a circular Gaussian with a FWHM of 1.8 arcseconds.
\label{fig5}}
\end{figure}
\clearpage

\begin{figure}
\vspace{19cm}
\includegraphics{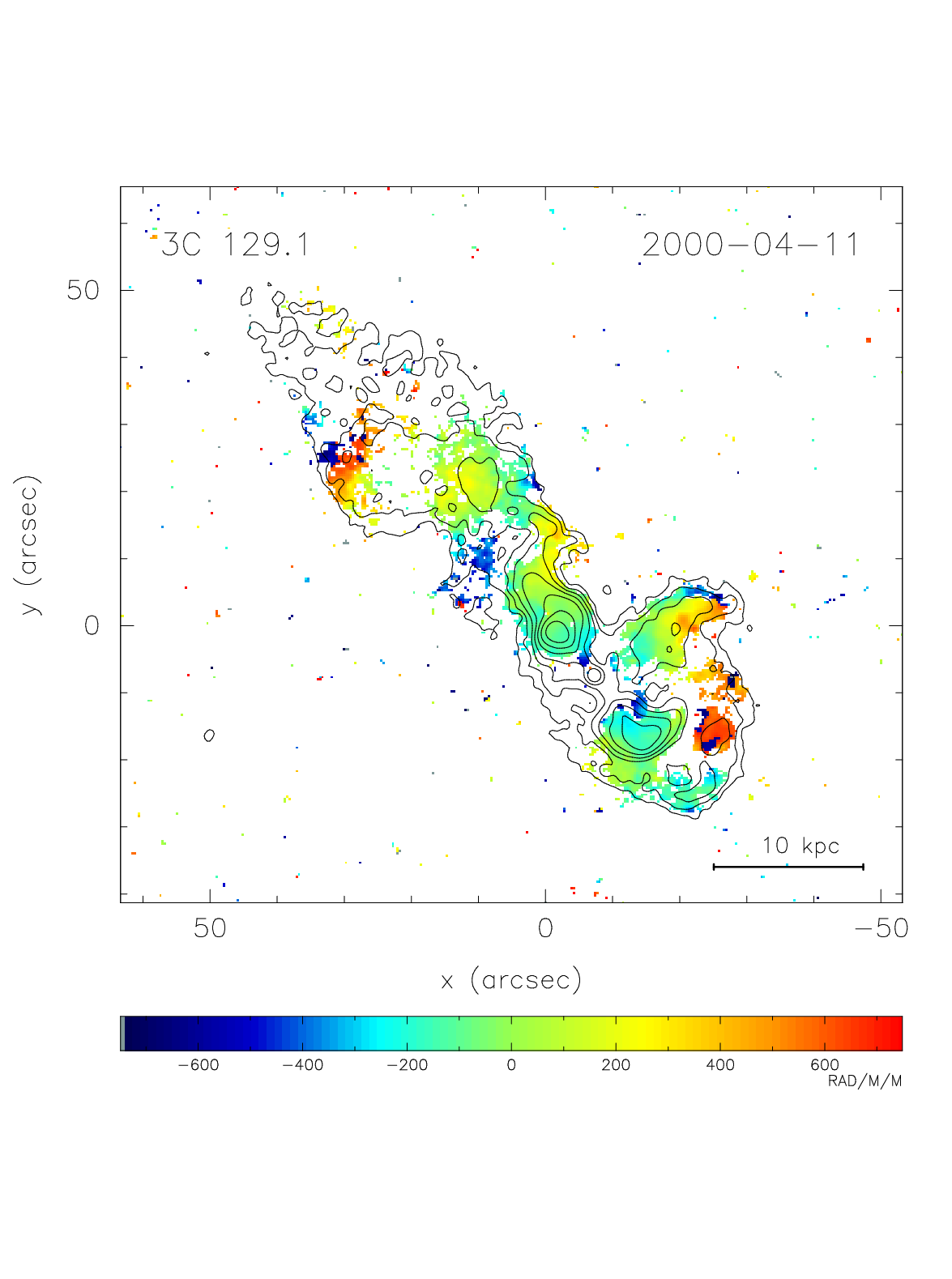}
\caption{Rotation measure image of 3C\,129.1 with contours of total
intensity at 4.9 GHz overlaid.   The colorbar ranges from $-$750 to $+$750
\radm.
\label{fig6}}
\end{figure}
\clearpage

\begin{figure}
\vspace{19cm}
\includegraphics{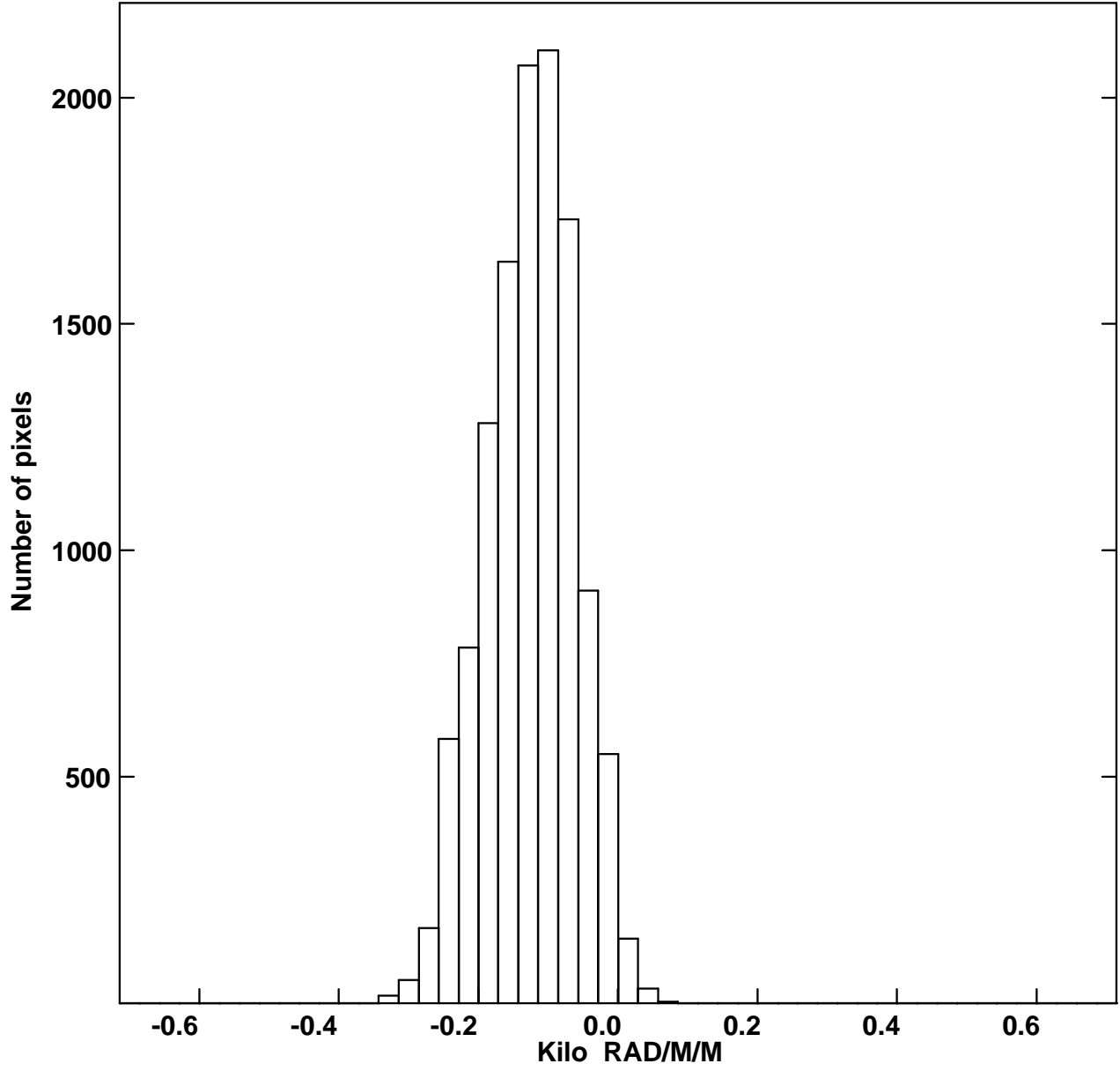}
\caption{Rotation measure distribution for 3C\,129.  The mean value
is $-$125 \radm\ with dispersion of 82 \radm.
\label{fig7}}
\end{figure}
\clearpage

\begin{figure}
\vspace{19cm}
\includegraphics{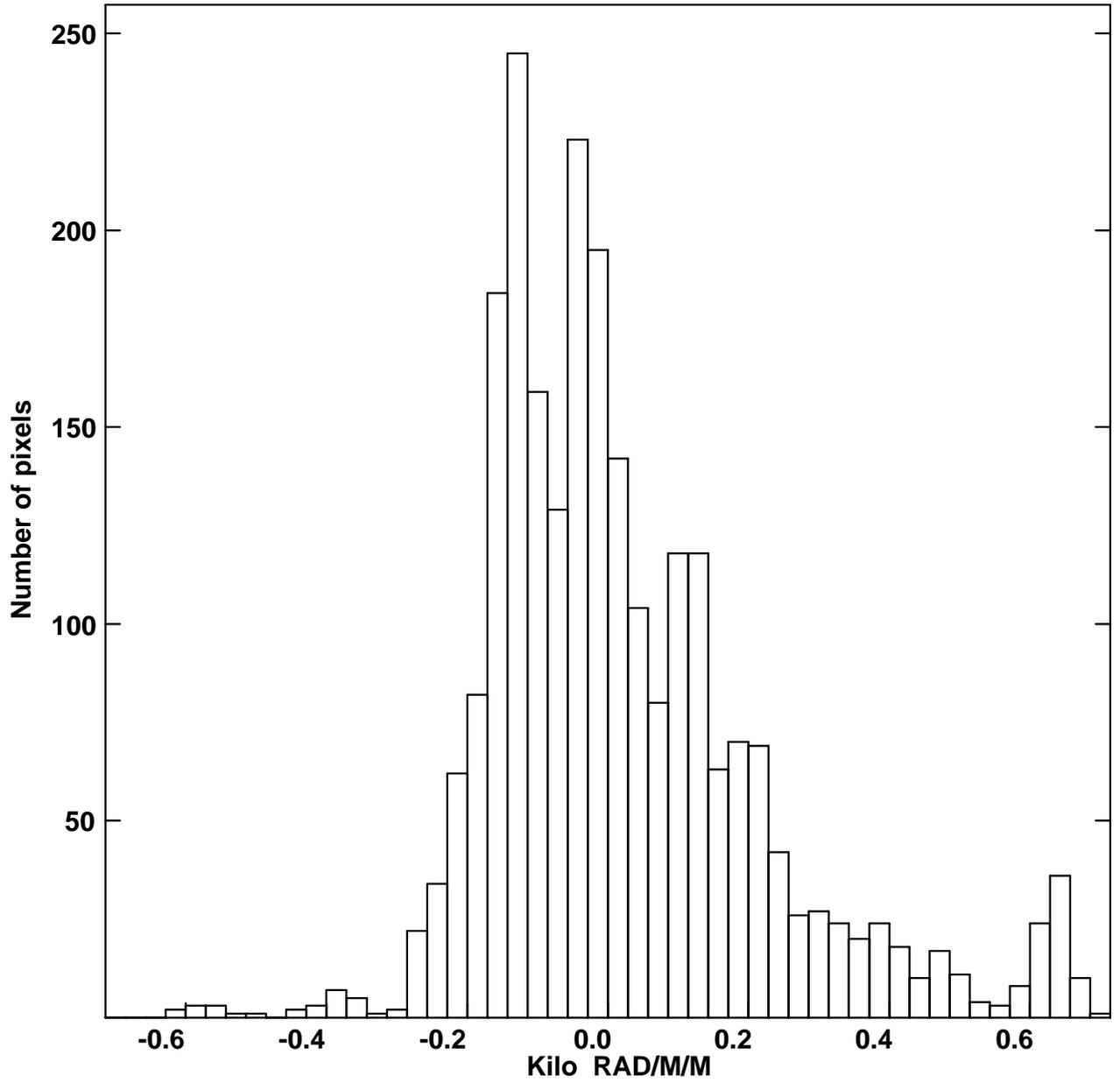}
\caption{Rotation measure distribution for 3C\,129.1. The mean value
is 21 \radm\ with dispersion of 200 \radm.
\label{fig8}}
\end{figure}
\clearpage

\def\dg{$^{\circ}$}
\begin{center}

TABLE 1 \\
\smallskip
O{\sc bservational} P{\sc arameters}
\smallskip

\begin{tabular}{l l r r r r r r r r}
\hline
\hline
Source & Date & Frequency & Bandwidth$^a$ & Config. & Duration \\
 &  & (MHz) & (MHz) &  & (hours) \\
\hline
\noalign{\vskip2pt}
3C\,129    & Jul1994 & 4585/4885 & 50 & B & 1.09 \\
         & Jul1994 & 7815/8165 & 50 & B & 1.06 \\
         & Jul1994 & 8515/8885 & 50 & B & 1.08 \\
         & Nov1994 & 4585/4885 & 50 & C & 0.21 \\
         & Nov1994 & 7815/8165 & 50 & C & 0.64 \\
         & Nov1994 & 8515/8885 & 50 & C & 0.60 \\
3C\,129.1  & Feb2000 & 4585$^a$/4885 & 50 & B & 1.75 \\
         & Feb2000 & 7815/8165 & 50 & B & 1.69 \\
         & Feb2000 & 8515/8885 & 50 & B & 1.79 \\
         & Apr2000 & 4585$^a$/4885 & 50 & C & 0.72 \\
         & Apr2000 & 7815/8165 & 50 & C & 0.90 \\
         & Apr2000 & 8515/8885 & 50 & C & 0.77 \\
\hline
\end{tabular}
\end{center}
$^a$ Due to an incorrect delay setting for the BD IF pair in the
5 GHz (6 cm) band, the 4585 MHz polarimetry from the Feb2000 and
Apr2000 runs were rendered unusable.

\clearpage

\begin{center}
TABLE 2 \\
\smallskip
S{\sc ource} P{\sc roperties} \\
\smallskip
\begin{tabular}{l c c}
\hline
\hline
Property & 3C\,129 & 3C\,129.1 \\
\hline
\noalign{\vskip2pt}
core RA (J2000) & 04$^h$49$^m$09\rlap{$^s$}{.\,}0760  & 04$^h$50$^m$06\rlap{$^s$}{.\,}6308 \\ 
\phantom{Core}Dec. (J2000) &  45\arcdeg 00\arcmin 39\rlap{\arcsec}{.\,}215 &  45\arcdeg 03\arcmin 05\rlap{\arcsec}{.\,}990 \\ 
radial velocity & 6240 km s$^{-1}$ & 6675 km s$^{-1}$ \\
projected distance  &  0.4 Mpc  & 0 Mpc \\
angular size & 20\arcmin & 1.5\arcmin \\
physical size & 500 kpc & 40 kpc \\
flux density (5 GHz) & 2650$^a$ mJy & 220 mJy \\
power (5 GHz) & 2.96 $\times$ 10$^{24}$ W Hz$^{-1}$ &  0.25 $\times$ 10$^{24}$ W Hz$^{-1}$ \\
$<$RM$>$ & $-$125 \radm & 21 \radm \\
$\sigma_{\rm RM}$ &  82 \radm & 200 \radm \\
$|$RM$_{\rm max}|$ & 260 \radm & 640 \radm \\
\hline
\end{tabular}
\end{center}

$^a$ Based on observations by Feretti \etal\ (1998) with the 
Effelsberg 100 m antenna.

\clearpage
\end{document}